\documentclass[]{article}
\usepackage{amsmath}
\usepackage{amssymb}
\usepackage{graphicx}
\usepackage[margin=1.0in]{geometry}
\usepackage[round]{natbib}

\newcommand{\captionfonts}{\small}

\makeatletter  % Allow the use of @ in command names
\long\def\@makecaption#1#2{%
  \vskip\abovecaptionskip
  \sbox\@tempboxa{{\captionfonts #1: #2}}%
  \ifdim \wd\@tempboxa >\hsize
    {\captionfonts #1: #2\par}
  \else
    \hbox to\hsize{\hfil\box\@tempboxa\hfil}%
  \fi
  \vskip\belowcaptionskip}
\makeatother   % Cancel the effect of \makeatletter

\title{General Equilibrium as a Topological Field Theory}
\author{Eric Kemp-Benedict\\
Stockholm Environment Institute\\
eric.kemp-benedict@sei-international.org}

\begin{document}

\maketitle
\bibliographystyle{plainnat}

\begin{abstract}
General equilibrium is the dominant theoretical framework for economic policy analysis at the level of the whole economy. In practice, general equilibrium treats economies as being always in equilibrium, albeit in a sequence of equilibria as driven by external changes in parameters. This view is sometimes defended on the grounds that internal dynamics are fast, while external changes are slow, so that the economy can be viewed as adjusting instantaneously to any changed conditions. However, the argument has not been presented in a rigorous way. In this paper we show that when conditions are such that: a) economies do respond essentially instantaneously to external influences; b) the external changes are small compared to the values that characterize the economy; and c) the economy's dynamics are continuous and first-order in time (as for Walrasian t\^atonnement), the resulting economic theory is equivalent to a topological field theory. Because it is a topological theory it has no dynamics in a strict sense, and so perturbatively---that is, when examining dynamics in the region of a critical point---the field theory behaves as general equilibrium posits. However, the field-theoretic form of the theory admits non-perturbative instanton solutions that link different critical points. Thus, in this theory, and in contrast to general equilibrium, the internal dynamics of the model occasionally make an appearance in the form of abrupt, noise-driven transitions between critical points.

\vspace{2em}

\end{abstract}

\section{Introduction}
Computable general equilibrium (CGE) models are the \textit{de facto} standard for economy-wide economic policy analysis. Despite their popularity, the theoretical foundation for general equilibrium is remarkably thin \citep{ackerman_still_2001,kirman_intrinsic_1989}. Theoretical support emerged in the search for ``microfoundations'' of macroeconomics carried out throughout much of the last century \citep{janssen_microfoundations:_1993}. That search ended in disappointment---the theory could not guarantee stability, so practitioners focused on finding equilibria, or critical points, regardless of stability \citep{ginsburgh_structure_2002}. The result is, in practice, equilibrium models with no dynamics that can take the system to equilibrium; instead, ``dynamics'' are provided by sequential equilibria that result from external changes---''shocks'' to the system.\footnote{The somewhat misleadingly-named theory of dynamic general equilibrium describes the sequence of equilibria that results from agents who optimize the use of their resources over some time horizon \citep{heer_dynamic_2009}.} These are represented in the model as exogenous changes in parameters, and so we refer to this as ``parametric'' dynamics.

In many presentations of general equilibrium the problem of how the equilibrium is reached---which we refer to as ``internal'' dynamics---is simply not addressed \citep{fisher_disequilibrium_1983,kirman_demand_2006}. Indeed, some presentations of general equilibrium argue that it is incorrect to think of an equilibrium as having been reached through some dynamic process \citep{ginsburgh_structure_2002}. We agree with \citet{fisher_disequilibrium_1983} that it is meaningless to talk about equilibrium without some notion of disequilibrium dynamics, and consider instead the justification sometimes offered in defense of general equilibrium that markets adjust rapidly enough to changed conditions that internal dynamics can be ignored compared to slow external changes. [E.g., see \citet{tobin_price_1993}, on the neoclassical assumption that markets clear instantaneously, a position he rejects.] In this paper we present a formal statement of this informal argument to assess whether it provides a justification for general equilibrium. Drawing on recent work by \citet{ovchinnikov_self-organized_2011,ovchinnikov_topological_2012}, we argue that when an economy can be considered as adjusting rapidly to slow, low amplitude noise, and the adjustment mechanism is first-order in time, it can be represented as a topological field theory. Under these conditions the theory behaves perturbatively as general equilibrium expects it to---it is a topological theory and there are, in a strong sense, no dynamics. However, the field theory also has non-perturbative instanton solutions that can move the system abruptly from one critical point to another. This can occur even without parametric dynamics---it occurs purely because of the internal dynamics in the presence of noise.

Because there is no dynamics in the topological theory, we drop the term ``equilibrium'' that is used in economics to refer to stationary solutions to the noise-free dynamics; instead, we refer to them as critical points. Instanton-mediated transitions occur only between critical points that can be reached ``classically'', in a sense that we clarify later in this paper. However, Ovchinnikov's theory implies, in the context of general equilibrium, that this will only happen when there are asymmetries in the response of prices or quantities of one set of goods with respect to another set of goods. Otherwise, a version of Anderson localization will keep the system at the critical point. In other words, the neglect, in general equilibrium, of the stability of equilibria may be justified in the limited case when an economy is driven by slow, low-amplitude noise, but not if there are asymmetric responses to changes in one set of goods compared to other goods; in this case, there will be occasional transitions from one critical point to another, mediated by instantons.

\section{Deterministic Theory}
To keep close to the theory of general equilibrium, we assume that quantities are fixed, and only prices can adjust, although this assumption is not essential for the theory presented in this paper. Adjustment is first order in prices, corresponding to a Walrasian adjustment process.\footnote{Walras himself did not present his theory as a dynamic equation; this formulation was proposed by \citet{samuelson_foundations_1947}. Also, we point out that adjustment mechanisms have been proposed that do not assume a t\^atonnement process \citep{fisher_disequilibrium_1983}, that include both prices and quantities \citep{flaschel_dynamic_1997}, and that are second-order in time \citep{kemp-benedict_price_2012}.} We define a vector of prices $p$, whose elements $p_1, p_2,\ldots,p_n$ are the prices of the $n$ goods that the economy produces. Prices adjust in the direction of excess demand $A(p,\pi)$ (the difference between demand and supply at the specified price), where $\pi$ is a vector of $m$ parameters and $A$ is a vector-valued function, with an element for each of the $n$ goods in the economy. This dynamic implies that when demand exceeds supply the price rises, and when supply exceeds demand the price falls.\footnote{Walras proposed that an ``auctioneer'' announces a public price, after which producers and consumers announce the quantities that they propose to sell or buy \citep{walras_elements_1954}. This is clearly unrealistic, and critics of general equilibrium often point to this flaw in the theory, asking who sets prices. In this paper we are not concerned with how prices are set, and accept the dynamic equation at face value.} That is,
\begin{equation}\label{eqn:first_order_dynamics}
\frac{dp}{dt} = A(p,\pi).
\end{equation}
Such dynamics do not guarantee stability in discrete time, even when the corresponding continuous-time model is stable \citep{saari_iterative_1985}, and it can be argued that economic processes necessarily take place in discrete time. However, the instabilities arise in models in which all prices are changed simultaneously and publicly. This is an unrealistic representation of real markets. \citet{gintis_dynamics_2007} provides a more realistic model in which prices are private, and, in an agent-based implementation of the model, finds that the system exhibits global convergence with local fluctuations. We proceed on the assumption that the continuous-time model is a good approximation to the actual process of adjustment in an economy, but is accompanied by ``noise'' that represents the difference between the approximation and the reality.

\subsection{Parametric dynamics}
A typical general equilibrium policy analysis first sets $dp/dt$ to zero in Equation (\ref{eqn:first_order_dynamics}) in both a ``no policy'' and a ``policy'' case; next, the critical points of $A(p,\pi)$ are found for fixed $\pi$, and the results compared. The policy is represented by setting one element of the parameter vector, e.g., $\pi_1$, equal to two separate values, $\pi_1$ (the no-policy value) and $\pi_1'$ (the policy value), where the path from $\pi_1$ to $\pi_1'$ is not specified.\footnote{Although we do not consider path-dependence of parametric variables in this paper, we note the problems that arise if multiple critical points appear after the introduction of the policy, as the standard method gives no way to choose between them. Common experience suggests that responses to a new economic policy depend on the path taken for the implementation. For example, an explanation of economic reforms in India by the Deputy Director of the Reserve Bank of India \citep{mohan_economic_2006} emphasized phased implementation of new policies. If the path does not matter for the outcome then it is difficult to see why the government would expend the effort on a phased implementation.}

\subsection{Classical paths}
From this point forward we focus on non-parametric dynamics, and so we suppress the parameter vector $\pi$ in subsequent equations. We note that Equation (\ref{eqn:first_order_dynamics}) is in a ``classical'' form, in that there is no noise. In contrast, later in this paper we discuss the case with noise, which is analogous to a quantum field theory. We therefore use ``classical'' in the sense that it is used in physics, rather than economics---the model we describe is a neo-classical economic model.

The excess demand function $A(p)$ is a vector-valued function. We assume that it is smooth,\footnote{This assumption is common, but questionable. We can expect continuity \citep{kirman_intrinsic_1989}, but not necessarily smoothness. \citet{ovchinnikov_topological_2012} discusses the case where $A(p)$ is non-differentiable, in particular if its is a fractal strange attractor.} and from the Helmholtz-Hodge theorem we can write it as the sum of the negative gradient of a scalar potential $V(p)$ and a divergence-free term $\bar{A}(p)$, so that
\begin{equation}\label{eqn:classical_dynamics_helmoltz_hodge}
\frac{dp}{dt} = -\nabla V(p) + \bar{A}(p).
\end{equation}
We dot-multiply both sides of this equation by $dp/dt$ and integrate over time, to find
\begin{equation}
\int_{t_0}^{t_f} dt\, \left|\frac{dp}{dt}\right|^2 = V\left(p(t_0)\right) - V\left(p(t_f)\right) + \int_{p_{\mathrm{cl}}} dp\cdot\bar{A},
\end{equation}
where $p_{\mathrm{cl}}$ is the classical path between $p(t_0)$ and $p(t_f)$. Because the integrand on the left-hand side of this equation is always positive, the right-hand side must also be positive. We therefore find, for classical paths, the condition
\begin{equation}\label{eqn:positivity_constraint}
V(p(t_f)) \le V(p(t_0)) + \int_{p_{\mathrm{cl}}} dp\cdot\bar{A}.
\end{equation}
This positivity condition continues to hold true in the presence of noise, but the ``classical'' path $p_{\mathrm{cl}}$ will be a tunneling path in that noise can assist it over a classical barrier, so that prices appear to tunnel through the barrier. We note that the path-dependent integral over $\bar{A}(p)$ means that the classical accessibility of a critical point depends on the path taken to reach it. However, this is not a serious issue for the classical theory, as our interest with this formula is whether the final price vector can be reached from the initial point; all that matters is that there is some path $p_{\mathrm{cl}}$ satisfying Equation (\ref{eqn:positivity_constraint}). Also, the path-dependent term in the positivity condition makes sense: it makes a path more favorable when it is positive, which will be true if the direction of change of a price along a path  is in the same direction as the spontaneous change that would occur if the excess demand were $\bar{A}$. We illustrate the reasonableness of the condition through an example in Appendix \ref{cha:appendix_potential}. In contrast to the classical case, the path-dependency of the second term matters a great deal for the field theory, where it determines whether the system remains localized at its critical points or undergoes a transition.

\section{Field Theory}
The ``classical'' theory presented in the previous section is deterministic, but in real economies we expect some deviation from the theoretical ideal. First, some prices may be influenced by external variations over which the market has no direct control (e.g., changing food prices from fluctuating crop yields or price movements on the international market). Second, the theoretical ideal requires all market actors to know the current actual state of excess demand and to respond appropriately; more realistically, market actors have to guess on the basis of incomplete information. Either of these situations can be represented in the model by introducing a noise vector $\xi$ with one entry for each good,
\begin{equation}\label{eqn:field_theory_dynamics}
\frac{dp}{dt} - A(p) = \xi.
\end{equation}
We further assume that changes in the noise fields are slow compared to the characteristic adjustment time of the price dynamics and also that the system provides no feedback to the noise fields (otherwise, they would be part of the system). The time scale separation between the noise and prices is what allows us to continue to write the excess demand as an instantaneous function of $p$ (that is, with no memory of earlier values of $p$). We recognize that this may be a questionable assumption in real economies. For example, manufacturers set prices infrequently, and based on criteria other than an assessment of the gap between current supply and estimated demand \citep{blinder_sticky_1994,blinder_asking_1998}, while retailers may change prices more frequently than manufacturers, but using a variety of heuristic strategies \citep{bolton_empirically_2003}. Also, government policies can influence price dynamics, as well as price levels. In a study of the reliability of a computable general equilibrium (CGE) model with an agricultural sector, the model was found to over-predict price volatility in importing countries; the authors proposed that government policies reduced the transmission of volatility from international to domestic markets \citep{valenzuela_assessing_2007}. However, in this paper we are concerned with a common justification for the assumptions of general equilibrium theory---that market adjustment processes (deterministic dynamics) are fast compared to the characteristic time of changes in exogenous factors (noise). We ask whether this assumption, if true, supports the conclusion that markets always stay close to the critical points of the excess demand function. For now, we note that the assumption may be false, and return to this point at the end of the paper.

The system described by Equation (\ref{eqn:field_theory_dynamics}) was discussed by \citet{ovchinnikov_self-organized_2011,ovchinnikov_topological_2012}. We present a simplified (and somewhat heuristic) derivation of the central result that the model described by Equation (\ref{eqn:field_theory_dynamics}) can be expressed as a topological field theory. From this point forward we will refer to the price and noise vectors as ``fields'', although we assume (with general equilibrium theory) that there is no spatial dependence.

The only stochastic elements in the model are in the noise, so the only statistical partition function we can construct is the one for the noise fields. We assume that the noise is Gaussian white noise with a noise-noise correlator $G_{ij}$,
\begin{equation}
\langle \xi_i(t)\xi_j(t') \rangle = G_{ij}\delta(t-t').
\end{equation}
With this assumption, the partition function $Z$ is
\begin{equation}\label{eqn:partition_fcn}
Z = \int \mathcal{D}\xi\, e^{-\frac{1}{2}\int_0^T dt\,\xi\cdot G\cdot\xi}.
\end{equation}
This partition function contains very little information and no dynamics, because the noise fields have a simple, and static, Gaussian structure. In the context of a path integral, however, Equation (\ref{eqn:field_theory_dynamics}) allows us to change variables from the noise fields $\xi$ to the price fields $p$. In general the mapping will be one-to-many, which has interesting implications for the behavior of the system. We write Equation (\ref{eqn:partition_fcn}) as
\begin{equation}\label{eqn:partition_fcn_varphi}
Z = \int \mathcal{D}p\, \left|\frac{\partial\xi}{\partial p}\right| e^{-\frac{1}{2}\int_0^T dt\,K\cdot G\cdot K},
\end{equation}
where
\begin{equation}
K = \frac{dp}{dt} - A
\end{equation}and $|\partial\xi/\partial p|$ is the determinant of the Jacobian matrix $J$, with elements
\begin{equation}
J_{ij} = \delta_{ij} \frac{d}{d t} - \frac{\partial A_i}{\partial p_j}.
\end{equation}
Using the standard technique of Faddeev-Popov ghosts, the Jacobian matrix can be expressed as the integral of the exponential of the Jacobian, bracketed by two vectors of Grassman variables $\eta$ and $\bar\eta$,
\begin{subequations}
\begin{align}
\left|\frac{\partial\xi}{\partial\varphi}\right| &= \det{J}\\
&= \int\mathcal{D}\eta\mathcal{D}\bar\eta\, e^{\int_0^T dt\,\bar\eta J \eta}\\
&= \int\mathcal{D}\eta\mathcal{D}\bar\eta\, e^{\int_0^T dt\,\left(\bar\eta\cdot d\eta/dt - \bar\eta\cdot\nabla A\cdot\eta\right)},
\end{align}
\end{subequations}
where $[\nabla A]_{ij} = \partial A_i/\partial p_j$. It is also convenient to re-express the integral over $p$ with the help of an additional bosonic field $B$, as
\begin{equation}
e^{-\int_0^T dt\,K\cdot G\cdot K} = M^{-1}\int\mathcal{D}B\, e^{-i\int_0^T dt\,B\cdot\left(K - \frac{1}{2}i G^{-1}\cdot B\right)},
\end{equation}
where $M$ is the Gaussian integral $\int\mathcal{D}B\,\exp(-B\cdot G^{-1}\cdot B)$. The equivalence can be checked formally by redefining $B_i$ as $\bar{B}_i - i \sum_j G_{ij}K_j$ and integrating over the $\bar B_i$ (for fixed $p_i$).
Putting all the pieces together, and ignoring the unimportant factor $M^{-1}$, the partition function becomes
\begin{equation}
Z = \int \mathcal{D}p\mathcal{D}B\mathcal{D}\eta\mathcal{D}\bar{\eta}\,e^{-S},
\end{equation}
where $S=\int_0^T dt\,L$ and the Lagrangian $L$ is
\begin{equation}
L = i B\cdot\left(\frac{dp}{dt} - A - \frac{1}{2}iG^{-1}B\right) - \bar{\eta}\cdot\left(\frac{d\eta}{dt} - \nabla A\cdot\eta\right).
\end{equation}

The topological nature of the theory emerges when we realize that the action can be generated from the (nilpotent) BRST operator
\begin{equation}
\mathcal{Q} = \int_0^T dt\, \left(\eta\cdot\frac{\delta}{\delta p} + i B\cdot\frac{\delta}{\delta\bar{\eta}}\right).
\end{equation}
The action $S$ can be expressed as
\begin{equation}
S = \lbrace\mathcal{Q},\Psi\rbrace,
\end{equation}
where
\begin{equation}
\Psi = \int_0^T dt\, \bar{\eta}\cdot\left(K - \frac{1}{2}i G^{-1}B\right).
\end{equation}
The action therefore consists solely of the BRST gauge-fixing term. This is consistent with the non-dynamical partition function in Equation (\ref{eqn:partition_fcn})---there were no dynamics in the beginning and therefore none at the end of the change of variables from $\xi$ to $p$. However, the nontrivial Jacobian that relates the variables allows topologically distinct regions of the space that the fields inhabit.

\subsection{Localization and instantons}
The symmetry under the $\mathcal{Q}$ operator  ($\mathcal{Q}$-symmetry) is responsible for the lack of dynamics in the perturbative field theory. As shown by \citet{ovchinnikov_self-organized_2011}, when the noise is low-amplitude ($|G_{ij}| \ll 1$) and the $\mathcal{Q}$-symmetry is unbroken, the partition function does not depend on the time $T$. At the one-loop level the partition function is a topological invariant, and as the ground state is $\mathcal{Q}$-invariant, higher-order perturbations will likewise be invariant. While the price field (with the conjugate bosonic field $B$) does, indeed, vary due to the noise \citep{ovchinnikov_topological_2012}, the variation is exactly compensated by fluctuations of the fermionic ghost fields, and the system is localized near the critical points of $A(p)$.

The $\mathcal{Q}$-symmetry is never broken if the excess demand function can be written purely as the gradient of a potential, so that $\bar{A}(p)$ in Equation (\ref{eqn:classical_dynamics_helmoltz_hodge}) is zero. Detailed arguments for this important claim are provided by \citet{ovchinnikov_self-organized_2011,ovchinnikov_topological_2012}, but we motivate it with a heuristic argument. Non-perturbatively, the field theory admits instanton solutions that link ground states. When the noise is very weak---as we assume in this theory---paths exceeding the barrier are strongly suppressed. However, if they are accompanied by anti-instantons that traverse the reverse path, then the barrier is effectively lowered \citep{levy_magnetism_2000}. As a consequence, in the low-noise case instantons are always associated with anti-instantons. In the theory presented in this paper, the instanton is a sequence of price changes leading from one critical point to another; the anti-instanton traverses the reverse sequence of price changes. If there are symmetrical price responses (so that $\nabla A$ is a symmetric matrix near the initial critical point), then the anti-instanton cancels the instanton, and the system is highly likely to remain at the initial critical point \citep{ovchinnikov_self-organized_2011,ovchinnikov_topological_2012}.

When the symmetry is unbroken, the fields change only because of the noise. The noise has no memory, and it drives the time-rate of change of the price field in Equation (\ref{eqn:field_theory_dynamics}), so changes in the price field are Markovian. When $A = 0$ and the noise is Gaussian, this corresponds to Brownian motion. When $A\ne 0$ and the amplitude of the noise is small, prices execute a random walk in the neighborhood of metastable critical points.

\subsection{Symmetry breaking and instanton-mediated transitions}
If the price response is not symmetric, so that $\bar{A}(p)\ne 0$, then the contribution of the instanton and the anti-instanton to the action are different. The difference arises from the path-dependency of the integral over $\bar{A}(p)$---the contribution of the reverse path (the anti-instanton) is not the same as that of the forward path (the instanton). Because the instanton and anti-instanton are imbalanced, the system is not localized at the critical point, and the system can transition from one critical point to another. In fact, it must transition through a path that joins critical points of different index (where the index of a critical point is the number of negative eigenvalues). The index of the critical point is its ``topological charge'', and the transition between topologically distinct states breaks the $\mathcal{Q}$-symmetry. Thus, to go from one stable point to another---that is, states with an index of zero---the system must pass through an unstable or saddle point \citep{ovchinnikov_self-organized_2011}.

This result implies that when there are asymmetric responses of the price of one good to changes in the price of another good, then the economy can, under the influence of slow and low-amplitude noise, occasionally and abruptly transition from one critical point to another. The transition is not deterministic, as it depends on the noise, so the timing of the transition cannot be known. However, the possibility for a transition can be known, if the shape of the excess demand function is known: a transition can occur whenever there is asymmetry in the way prices of goods respond to one another, and the inequality (\ref{eqn:positivity_constraint}) is satisfied.

\section{Implications for CGE theory}
Reviewing the arguments made by CGE proponents for their practices, we find that in the case where the system's dynamics are fast compared to the slow, low-amplitude noise, two of those arguments seem to be at least partly valid:
\begin{enumerate}
\item there are no dynamics in the theory (beyond instanton-mediated transitions),
\item the system can live at an metastable critical point (but only if the excess demand function can be expressed as the negative gradient of a scalar potential).
\end{enumerate}
However, there is no reason to think that the excess demand function will have the second property, and so we add a further point,
\begin{enumerate}
\setcounter{enumi}{2}
\item if there are asymmetries in the way that quantities or prices of some goods respond to changes in quantities or prices of other goods, then the system will rarely but occasionally transition from one critical point to another.
\end{enumerate}

It is important to recognize that the assumption of slow and low-amplitude noise is often not realistic. For example, a small, open, agricultural economy will be strongly influenced by international prices for the commodities that it produces. Those commodity prices can vary considerably during a growing season, so that by the time the ``system'' (composed of, e.g., farmers and a commodity board) is able to respond, the ``noise'' (international commodity prices) may have changed many times and with a large amplitude. Therefore this partial justification for general equilibrium theory should be interpreted with caution. It is only relevant when an economy is subject to slow and low-amplitude fluctuations. Moreover, the validity of the assumption depends on the critical points being locally stable against small perturbations, and current economic theory cannot say what economic conditions produce stable critical points.

These (important) cautions aside, if an economy is well-described by the theory presented in this paper, then it can still undergo quite dramatic and rapid shifts. But these shifts are the instanton-mediated transitions from one critical point to another; after the transition the system is once again characterized by fast dynamics and slow, low-amplitude noise.

\section{Conclusion}
The practice of computable general equilibrium (CGE) modeling dominates in economy-wide policy analysis, yet its theoretical foundation is weak. We have argued that partial support for the theory of general equilibrium comes from a surprising direction---quantum field theory. The connection was elucidated in a general way by \citet{ovchinnikov_topological_2012}, while the present paper applies his theory to macroeconomics. The implications of the theory are that, if the assumption that an economy can be split into two parts---a complex dynamic part that responds very quickly to a slow, low-amplitude part that can be regarded as stochastic noise---then the economy can feature abrupt transitions. It will spend most of its time at a single critical point, but if there are asymmetric responses to changes in prices of different goods, then the system will occasionally transition from one critical point to another. Because instantons are non-perturbative phenomena, they do not emerge in perturbative macroeconomic models.

In this paper we have followed standard general equilibrium theory and assumed no spatial dependence. In the standard theory of general equilibrium this is justified by labeling commodities by their location and time, so that a sack of kidney beans in Paris, Texas today is a completely different commodity than a sack of kidney beans in Paris, France tomorrow \citep{ginsburgh_structure_2002}. This may be reasonable when considering each of the markets independently, but not as a whole---the global collection of local markets in kidney beans is the global market in kidney beans. This suggests that a spatially explicit representation of interconnected markets (perhaps as a network) could reveal some interesting dynamics. Indeed, \citet{ovchinnikov_self-organized_2011,ovchinnikov_topological_2012} points out that a spatial theory with a large number of critical points can exhibit self-organized criticality. Such behavior could, for example, help to explain the scale-independent behavior of financial markets seen by \citet{preis_switching_2011}.

The theory presented in this paper gives a picture of an economy that is partly in agreement with those who defend the theory of general equilibrium, but partly at odds with it. Unlike the general equilibrium view, in which the economy can only change through the influence of a secular external change (a ``shock''), in the view proposed in the present paper the economy can change due to its own internal logic, through a spontaneous transition under the influence of slowly varying, low-amplitude noise.

\bibliography{cge_tft}

\newpage
\appendix
\section{Multiple Equilibria and Path-Dependency}\label{cha:appendix_potential}
\begin{figure}[htb]
\centering
\includegraphics[scale=0.75]{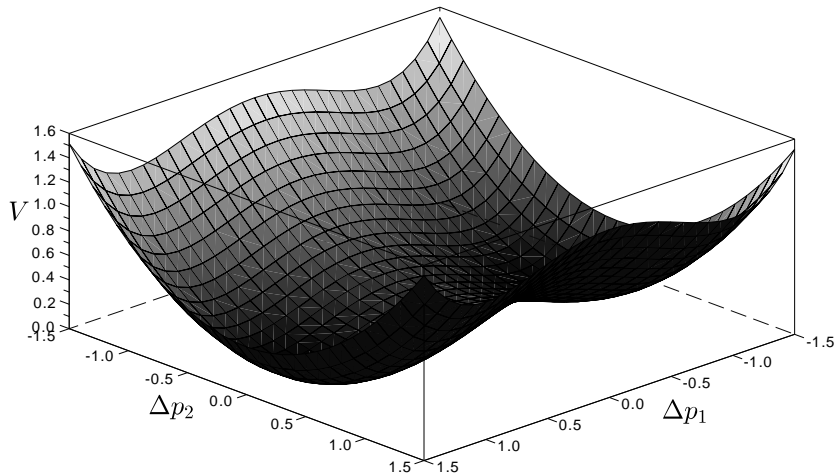}
\caption{Symmetric double well potential}
\label{fig:double_well}
\end{figure}

In this appendix we show how, in an example, the path-dependent term that appears in the positivity condition in Equation (\ref{eqn:positivity_constraint}) favors an economically sensible path. For our example we define an excess demand function that has two interesting features: multiple equilibria and asymmetric responses between different prices. We construct it in the following way. First, we define
\begin{subequations}
\begin{align}
\frac{dp_1}{dt} &= -\frac{\partial}{\partial p_1}V(p_1, p_2) + k (p_2 - p_2^*),\\
\frac{dp_2}{dt} &= -\frac{\partial}{\partial p_2}V(p_1, p_2) - k (p_1 - p_1^*).
\end{align}
\end{subequations}
The terms multiplied by $k$ are asymmetrical, and lead to path dependency. In this equation, $(p_1^*,p_2^*)$ is a reference point for the prices, allowing us to work with the deviations $(\Delta p_1,\Delta p_2)$, where
\begin{equation}
p_i = p_i^* + \Delta p_i,\ i=1,2.
\end{equation}
We construct a symmetric double-well potential with two minima at the same level. The general form for the potential is
\begin{equation}
V(p_1, p_2) = V(p_1^*+\Delta p_1, p_2^*+\Delta p_2) = \frac{1}{4}\left(a^2 - \Delta p_1^2\right)^2 + \frac{1}{2}b \Delta p_2^2.
\end{equation}
A specific example, for $a=1$, $b=1$, is shown in Figure \ref{fig:double_well}.

With this potential we have
\begin{subequations}\label{eqn:appendix_dynamics}
\begin{align}
\frac{d\Delta p_1}{dt} &= \Delta p_1\left(a^2 - \Delta p_1^2\right) + k\Delta p_2,\label{eqn:appendix_dynamics_a}\\
\frac{d\Delta p_2}{dt} &= -b \Delta p_2 - k\Delta p_1.\label{eqn:appendix_dynamics_b}
\end{align}
\end{subequations}
We now tell a story about this system to motivate its dynamics. Considering only the terms multiplied by $k$, assume that Good 1 is a manufactured item that requires Good 2 for its manufacture. If the price of Good 2 goes up, then manufacturers of Good 1 eventually raise their prices to compensate, so the coefficient of the price of Good 2 in Equation (\ref{eqn:appendix_dynamics_a}) is positive. In contrast, if the price of Good 1 goes up, then fewer people buy Good 1 and therefore demand for Good 2 (the input) goes down. This leads to a fall in its price, so the coefficient of the price of Good 1 in Equation (\ref{eqn:appendix_dynamics_b}) is negative. Good 2 has a standard own-price response that restores the system to equilibrium: if sellers put the price up, then demand falls and they respond by lowering the price again. For Good 1, the own-price response is more complicated. In the vicinity of the two minima, the system exhibits a standard own-price response of restoring the price to its equilibrium location. However, there are two possible minima, one at a lower price than the other. We suppose in this case that there are two markets that manufacturers can potentially sell into: a larger but low-spending group looking for a bargain and a smaller but high-spending group that values uniqueness---they will pay a premium for goods that few people have.\footnote{Standard consumption theory claims to not question people's preferences, and instead discovers them through their consumption behavior. We could therefore make almost any story we like to motivate this example; indeed, almost any continuous function can be represented as the sum of well-behaved individual demand functions \citep{kirman_intrinsic_1989}, so we do not need any motivation at all. However, this particular story---that people value having something others do not---is rarely if ever applied in conventional economic models. In its defense, the story we tell here draws on the concept of ``positional goods'' proposed by \citet{hirsch_social_1978}, and we note that there is empirical evidence of positional spending \citep{carlsson_you_2007}. Another explanation for the same excess demand function is that consumers associate price with quality \citep{gabor_price_1966}, a phenomenon which is, like positional consumption, problematic for conventional economics \citep{stiglitz_causes_1987}. Moreover, consumers' beliefs are not always justified, because for many goods the relationship between price and quality is weak \citep{curry_prices_1988,gerstner_higher_1985}, so essentially the same good can be sold at a higher price if it is believed to be superior.} If manufacturers are, for example, producing for the low-spending group, then they find it hard to move away from that stable point, because they cannot raise prices too much for the low-spending group, but cannot induce the high-spending group to buy.

\begin{figure}[htb]
\centering
\includegraphics[scale=0.5]{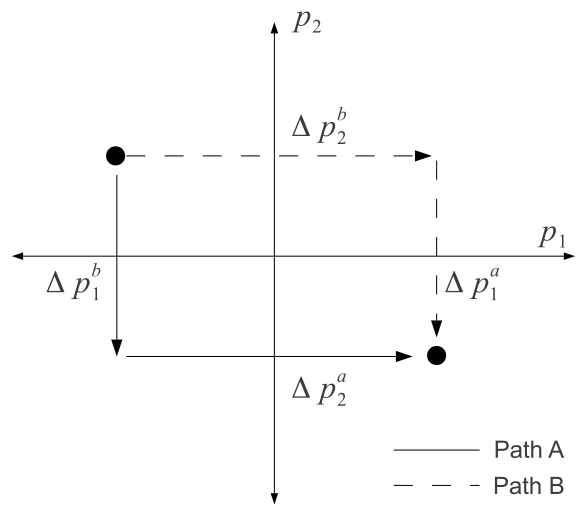}
\caption{Paths from one stable critical point to another}
\label{fig:price_paths}
\end{figure}

Setting the time-derivatives of $\Delta p_1$ and $\Delta p_2$ to zero shows the conditions for the two wells to be
\begin{subequations}
\begin{align}
\Delta p_1 &= \pm\sqrt{a^2 - \frac{k^2}{b}},\\
\Delta p_2 &= -\frac{k}{b} \Delta p_1.
\end{align}
\end{subequations}
So there are two stable equilibrium points, $(\Delta p_1^b,\Delta p_2^b)$ and $(\Delta p_1^a,\Delta p_2^a)$, as shown in Figure \ref{fig:price_paths}.

We now suppose that the system passes from $(\Delta p_1^b,\Delta p_2^b)$ to $(\Delta p_1^a,\Delta p_2^a)$, following one of two paths, as shown in Figure \ref{fig:price_paths}. In Path A, the price of Good 2 first falls, and then the price of Good 1 increases. In Path B, the price of Good 1 increases, and then the price of Good 2 falls. For Path A,
\begin{equation}
\int_A dp\cdot\bar{A} = -k\Delta p_1^b \left(\Delta p_2^a - \Delta p_2^b\right) + k\Delta p_2^a \left(\Delta p_1^a - \Delta p_1^b\right) < 0,
\end{equation}
while for Path B,
\begin{equation}
\int_A dp\cdot\bar{A} = k\Delta p_2^b \left(\Delta p_1^a - \Delta p_1^b\right) - k\Delta p_1^b \left(\Delta p_2^a - \Delta p_2^b\right) > 0.
\end{equation}
Examining Equation (\ref{eqn:positivity_constraint}), we see that the path that gives a positive value for this integral is favored. Therefore, Path B is favored over Path A. This is reasonable, because, for Path B to come about, manufacturers switch to the higher price that drives the bulk of consumers away but that attracts a small client\`ele that is willing to pay for uniqueness. However, if the amount of the input, Good 2, per unit of output, Good 1, remains the same, then demand for Good 2 is lower and the price falls. That is, Path B has a sensible story, given the assumptions about consumer behavior. In contrast, Path A does not. For Path A to come about, first the price of Good 2 falls (for example, because new reserves are discovered, new producers enter the market, or other consumption of the good falls), and then manufacturers of Good 1 decide that it is a good time to move from the lower-paying to the higher-paying group of customers. They might decide to do so, but not because of the drop in the price of Good 2, which would otherwise give them a windfall profit while the drop persisted.

While this is a simple example, it illustrates the point that the the positivity condition in Equation (\ref{eqn:positivity_constraint}) selects for economically reasonable paths. It does this because it favors paths where the changes are in the direction that would occur spontaneously if the excess demand function were given by its non-potential component $\bar{A}$. If $\bar{A}$ itself is reasonable, in that it reflects a sensible causal story about the relationship between goods, then the favorable path will likewise reflect a sensible causal story.

\end{document}